\documentclass[12pt]{iopart}
\usepackage{graphicx}
\usepackage{dcolumn}
\usepackage{bm}
\usepackage{subfigure}
\usepackage{hyperref}
\usepackage{xcolor}
\usepackage[utf8]{inputenc}

\expandafter\let\csname equation*\endcsname=\relax
\expandafter\let\csname endequation*\endcsname=\relax
\usepackage{amsmath}
\usepackage{amssymb}

\DeclareUnicodeCharacter{2009}{\,} 
\DeclareUnicodeCharacter{03B2}{beta} 

\begin{document}

\title{On the stabilisation of locked tearing modes in ITER and other large tokamaks}

\author{R. Nies}
\ead{rnies@pppl.gov}
\author{A.H. Reiman}%
 \ead{reiman@pppl.gov}
\author{N.J. Fisch}
\ead{fisch@princeton.edu}
\address{%
 Princeton Plasma Physics Laboratory, Princeton University, Princeton, New Jersey 08544, USA
}%

\date{\today}

\begin{abstract}

Tearing modes in tokamaks typically rotate while small and then lock at a fixed location when larger. Research on present-day devices has focused almost exclusively on stabilisation of rotating modes, as it has been considered imperative to avoid locked modes. However, in larger devices, stabilisation during the rotating phase is made difficult by fast locking at small island widths, and large broadening of the stabilising wave-driven current profile. In contrast, the smaller island width at locking not only mitigates the deleterious consequences of locked modes, but also permits their efficient stabilisation. On large devices, it thus becomes surprisingly advantageous to allow the mode to grow and lock naturally before stabilising it, challenging the mainstream strategy of neoclassical tearing mode stabilisation during the rotating phase. Calculations indicate that a locked island stabilisation strategy should be adopted in the ITER tokamak, with a large potential impact on the fusion gain and disruptivity.
\end{abstract}

\maketitle

\section{Introduction}

While a high thermal pressure is desirable for tokamak performance, it can lead to the destabilisation of neoclassical tearing modes (NTMs), giving rise to magnetic islands which degrade confinement and cause disruptions. Typically, NTMs initially rotate with the background plasma, then decelerate due to the torque from interaction with a resistive wall surrounding the plasma, and finally lock to the machine error field. NTMs can be stabilised e.g. by using radio-frequency current drive (CD) \cite{reiman_suppression_1983}.

Although the positioning of locked islands \cite{volpe_advanced_2009} in front of the wave launchers and their subsequent stabilisation has been demonstrated experimentally \cite{volpe_avoiding_2015}, research to date has largely focused on the stabilisation of islands during the rotating phase. In this work, we investigate the possibility of locked island stabilisation in large tokamaks like ITER, and the circumstances under which this strategy would be preferable to that of rotating island stabilisation. In this scenario, we would allow the islands to lock to the ambient error field and then rapidly suppress the islands using RF driven currents, which can operate much more efficiently once the island is locked. This entails adjusting the error field correction coils such that the residual error field has the appropriate phase that places the O-point of the locked island in front of the launcher.  

The large size of ITER compared to present-day medium-sized devices is expected to lead to comparatively small rotation frequencies. When taking into account the effect of ITER's blanket modules, the $2/1$-NTM has been predicted to lock rapidly, at a small width of only $w_\mathrm{lock} \approx 9$~cm \cite{la_haye_effect_2017}, which corresponds to only $4.5\%$ of the minor radius $a$. Furthermore, a strong broadening of the radial current profile driven by electron cyclotron (EC) waves is expected in large devices like ITER. The edge density fluctuations in ITER are predicted to cause a broadening by a factor of $2.5$ to $3.5$ \cite{snicker_effect_2018}, significantly reducing the stabilisation efficiency of rotating island stabilisation schemes. These recent insights indicate that stabilisation of NTMs while they are rotating will be challenging in large tokamaks. They further suggest that ITER may already be in a regime where locked island stabilisation should be considered. 

As we will discuss, the issue of the appropriate adjustment of the phase of the residual error field to provide the desired position of the locked island is one that arises more generally. If an island locks for whatever reason, it will be desirable for its O-point to lock in front of the electron cyclotron wave launcher so that stabilisation via RF driven currents remains a viable option.

This paper challenges the currently accepted wisdom that neoclassical tearing modes must be stabilised whilst they are rotating. The potential advantages of adopting a locked island stabilisation strategy motivate a shift in the theoretical and experimental research agenda to bring about commercially viable tokamak fusion. 

The paper is organised as follows: in section~\ref{sec:concerns}, we address concerns about the impact of locking on the subsequent evolution of the plasma. As will be discussed, the predicted small width of the islands at locking helps to alleviate those concerns. Section~\ref{scenarios} will discuss the implications of recent findings indicating that stabilisation of NTMs during the rotating phase in large tokamaks such as ITER will be more difficult than had been earlier recognd. That section will also introduce the stabilisation scenarios that have been previously considered, which will be compared with our proposed locked island stabilisation scenario.

Sections \ref{sec:numerical_model} and \ref{sec:power} will compare numerical time-dependent simulations of the differing stabilisation scenarios, focusing particularly on parameters appropriate to the ITER Scenario 2 equilibrium. These sections will be followed by  section~\ref{sec:island_stab_large_wdep}, which will make use of approximate analytical solutions to provide a broader perspective, allowing comparison of power requirements for a broader set of parameters.

Section~\ref{sec:numerical_model} will describe the numerical model of the time evolution that we have used to explore the power requirements for each of the scenarios. Section~\ref{sec:power} will describe calculations of the peak power requirements making use of that model. It will be shown there that, for the parameter regime considered, the model predicts that stabilisation of the mode after it locks, instead of stabilisation during the rotating phase, would lead to reduced peak and average power requirements on the EC system.

In the context of the approximate analytical solutions, section~\ref{sec:ITER_error_field_requirements} will address the question of what constraints on the magnitude of the error field must be satisfied if locked mode stabilisation is to be attractive.  If the error field is too large, that can impact the power required to reduce the NTM locked island width below the marginal width for NTM stability. If the error field is too small, the island cannot be positioned in front of the wave launcher. It will be seen that, for ITER, the resulting constraint on the error field magnitude is comparable to that believed to be necessary in order to avoid the appearance of born locked modes.

Section \ref{sec:discussion} will discuss the results of the paper, and section~\ref{conclusions} will summarise our conclusions. We will argue that the adoption of a locked island stabilisation strategy could potentially have a large impact on the fusion gain in large devices like ITER, and could free up EC power for other purposes. Furthermore, even if a locked mode stabilisation strategy for NTMs is not adopted, locked mode stabilisation could assist in reducing the disruptivity by providing a back-up stabilisation scheme for those cases where rotating island stabilisation fails and mode locking occurs. This should motivate theoretical and experimental study of the requirements for successful locked mode stabilisation.

\section{Concerns about locking}
\label{sec:concerns}

Although the stabilisation of locked islands using wave-driven currents has been demonstrated experimentally \cite{volpe_avoiding_2015}, other experimental research (e.g. \cite{gantenbein_complete_2000, isayama_complete_2000, la_haye_control_2002}) has focused almost exclusively on island stabilisation during the rotating phase, before locking occurs. Similarly, theoretical calculations of island stabilisation for ITER have focused on stabilising the island during the rotating phase (e.g. \cite{poli_criteria_2015,poli_electron_2018}). This strategy has emerged from a concern about the damaging impact of locked islands on plasma performance, and in particular from a concern about the impact on disruptivity.

\subsection{Disruptivity}
\label{disruptivity}

A concern about locked modes has arisen, in part, from the fact that most disruptions are preceded by the growth of locked islands. It has been reported that 95\% of the disruptions on JET are preceded by the presence of such islands \cite{gerasimov_overview_2018}.  However, an analysis of the mode amplitudes at disruption found that locked modes in JET with the ITER-like wall triggered disruptions at a distinct mode amplitude corresponding to a large island width of about 30\% of the minor radius \cite{de_vries_scaling_2016}. In ITER, the $2/1$-NTM is predicted to lock at a small width of only $4.5\%$ of the minor radius \cite{la_haye_effect_2017}, see Section~\ref{sec:numerical_model}.  Magnetic islands, even locked magnetic islands, grow on a resistive time scale. There is a significant margin between the mode locking and disruption events. As an example, Figure 1 in Ref. \cite{de_vries_scaling_2016} shows the time evolution of a born locked mode in JET, for which the width of the locked island is initially very small. The island grows for 500 ms before it triggers a disruption.

\subsection{Acceleration of growth}
\label{growth}

Another factor that has led to a concern about locking is the observation that the growth of the island may accelerate after locking.  The resistive wall boundary condition is stabilising for rotating islands, but not for locked islands. Also, the resonant component of the error field is stabilising for the rotating islands and destabilising for the locked ones. These effects are particularly apparent in contemporary tokamaks, where the rotating islands may grow quite large before locking, and may approach or even reach their saturation widths, only to resume growth after locking.  Although an island may grow more rapidly after it locks, it nevertheless grows on a slow resistive time scale.

Ref. \cite{volpe_avoiding_2015} describes a set of experiments in DIII-D in which unstable islands were locked at relatively large width and then stabilised by electron cyclotron current drive (ECCD). Figure 1 in that paper shows two shots in which a resonant perturbation was applied to lock growing islands, with the shots otherwise identical except that in one of the shots ECCD was turned on to suppress the locked island.  In both shots an NTM grew for about 180 ms until the rotating island was locked by an applied perturbation when it reached a width of about 15\% of the minor radius. In the shot where the ECCD was not turned on, the island continued to grow for about another 650 ms after locking before it triggered a disruption when it reached a width of about 30\% of the minor radius. In the shot in which ECCD was turned on, the locked island was rapidly stabilised, and the plasma remained in H-mode. 
It should be noted that, once islands become sufficiently large to trigger a disruption, the subsequent time evolution can proceed quite rapidly. This is the case for both rotating and locked islands.  It only happens when the island width is quite large, and is not relevant for the small islands discussed in this paper.

\subsection{Loss of H-mode}
\label{H-mode}

Another concern relates to the loss of H-mode incurred after mode locking, as the locked island viscously brakes the background plasma rotation \cite{zohm_plasma_1990, snipes_plasma_1990}. As loss of H-mode is thought to proceed on a momentum confinement timescale \cite{nelson_experimental_2020} however, H-mode can be preserved by quickly stabilising the island after locking, as was experimentally demonstrated for large locked islands in DIII-D \cite{volpe_avoiding_2015, volpe_private_2021}. Furthermore, small locked islands might not have as deleterious an impact on the background rotation \cite{klevarova_validation_2020}, and the small locked NTMs considered here might not lead to loss of H-mode, similarly to the small locked modes induced by resonant magnetic perturbations (RMPs) in the context of edge localised mode (ELM) suppression \cite{evans_suppression_2004}.  There is a paucity of data on the effects of locking by islands of the size expected in ITER.  It is likely that ITER will provide such data, but it would be desirable to obtain such data in advance from existing devices in order to be prepared to optimise the utilisation of the valuable machine time on ITER.

\subsection{Nonaxisymmetric field required for locking at the desired phase}
\label{phase}

ITER will have a set of nonaxisymmetric coils whose purpose will be to compensate for the error fields that arise from the finite tolerances in the placement of the toroidal and poloidal field coils. A description of these coils can be found in Ref. \cite{amoskov_optimization_2015}. These coils are intended to compensate for $n=1$ field errors, and for that purpose they will produce resonant $n=1$ fields intended to cancel $(1,1)$, (2,1) and (3,1) Fourier components at the corresponding rational surfaces.  It is intended that the error compensation coils reduce these resonant error fields by a factor of about 4.

When rotating NTM islands in ITER are sufficiently slowed by their interaction with the resistive wall, they will lock to the residual error field that remains after partial compensation by the correction coils. We require only that the error compensation coils be adjusted such that the residual error field has the desired phase. No separate applied perturbation is required.

If ITER does attempt to stabilise rotating islands before they lock, this may not be successful 100\% of the time.  More generally, the experience on JET suggests that the great majority of disruptions in ITER will be preceded by the growth of locked islands, and that they will mostly arise from off-normal events other than NTMs \cite{gerasimov_overview_2018,de_vries_influence_2014}. When the islands lock, their phase will be determined by the phase of the residual error field.  If this phase is such that the O-point of the locked island aligns with that of the EC launcher, it may still be possible to suppress the island at that point and avert a disruption. That will not be possible if the phases do not align.

In any case, we believe that it is important that these issues be investigated on contemporary tokamaks.  The impact of locking for islands in the range of 4\% to 5\% of the minor radius has not been studied.  The issue with respect to the phase of the residual error field is likely to display itself in the early stages of ITER operation, before disruptions threaten damage to the device.  It would nevertheless be undesirable to waste valuable run time on ITER investigating the issue.

\section{Envisioned stabilisation scenarios and implications of recent findings}
\label{scenarios}

In ITER, NTM stabilisation is planned using EC waves through the Fisch-Boozer current drive effect \cite{fisch_creating_1980,fisch_theory_1987}. Studies have largely focused on the stabilisation of the islands produced by NTMs while they are still rotating. (e.g.~\cite{de_lazzari_merits_2009, sauter_requirements_2010, bertelli_requirements_2011, poli_electron_2018}). We will refer to this as rotating mode (RM) stabilisation.  The primary purpose of the ITER EC upper launcher is NTM stabilisation. It has been designed to have a fixed toroidal launching angle $\beta$ of about $20^\circ$ on the basis of ray tracing calculations for RM stabilisation via continuous EC injection \cite{ramponi_iter_2007}. More recent experimental evidence indicates that the deposition profile of the ECCD is significantly broadened relative to the predictions of ray tracing codes \cite{decker_effect_2012,brookman_experimental_2017, chellai_millimeter-wave_2018, chellai_millimeter-wave_2019,brookman_resolving_2021}. Theoretical calculations for ITER predict a broadening by a factor of 2.5 to 3.5 at the $q=2$ surface due to scattering  of the EC beam by edge density fluctuations \cite{snicker_effect_2018}. The power requirements for stabilisation via continuous EC become prohibitive for deposition profiles that are that broad  \cite{poli_recent_2015, snicker_effect_2018}. RM stabilisation remains a possibility only if the ECCD is modulated at the island rotation frequency, with the phase controlled such that the ECCD is deposited in the neighborhood of the island O-point.

In addition to the issue of broadening, islands in ITER are now predicted to lock more quickly than previously thought, and, as will be discussed below, this impacts the power requirements for RM stabilisation for both continuous and modulated RF. The revised estimate of the locking time has emerged from an analysis of the effect of the test blanket modules, which has also found a critical island width for slowing of about 5 cm \cite{la_haye_effect_2017}. Islands above the critical width will continue to slow until they lock.  Modulated RF can only be used when the island width is above the detection threshold, because it requires knowledge of the island phase.  One concern is that, if the critical island width is found to be less than the detection threshold in ITER, it will not be possible to use modulated ECCD to prevent the island from locking.

The reduced width at locking in ITER is a common feature of envisioned large tokamaks, due to the low plasma rotation. Indeed, compared to contemporary medium-sized devices where the rotation is mostly due to neutral beam injection, the inertia is substantially increased, while the torque is, at best, only marginally increased. In an ignited device, the neutral beam torque may be entirely absent. Although some intrinsic rotation may still be present, it is expected to remain modest in large tokamaks. A predicted scaling of the intrinsic rotation is $\omega_\mathrm{intr.} \sim c^2 T_e / e I R$, with the speed of light $c$, electron temperature $T_e$, elementary charge $e$, plasma current $I$ and major radius $R$ \cite{parra_scaling_2012}, generally leading to smaller intrinsic rotation frequencies for large tokamaks due to the increased $I$ and $R$.

A particularly serious drawback for modulated ECCD is that the threshold island width for triggering NTMs in ITER is expected to be smaller than the island detection threshold \cite{hender_chapter_2007}, and therefore smaller than the threshold width for the diagnostics to accurately follow the island phase.  That implies that modulated ECCD will not be capable of shrinking the island below the marginal stability width, and must remain on throughout the shot to prevent the island from growing, reducing  the fusion gain $Q$.  For the application of continuous, rather than modulated, ECCD, it has been argued that, even for significantly smaller deposition broadening factors than are currently being predicted, it is advantageous to apply power preemptively \cite{poli_electron_2018}. Again, having the power on throughout the shot would reduce $Q$.

Locked mode (LM) stabilisation has a potential advantage that it can be used to shrink the island below the marginal stability width, so that it can then be turned off until the next triggering event. This will be advantageous in terms of the impact on $Q$ if NTMs are not being continuously triggered. It is intended that ITER will operate ELM-free, and tokamak reactors will also likely need to operate without ELMs, which would eliminate them as an NTM trigger.  The other major trigger of NTMs in contemporary tokamaks is sawteeth. The time between sawteeth in ITER is predicted to be much larger than the energy confinement time \cite{hender_chapter_2007}.  If sawteeth are the primary trigger of NTMs in ITER, stabilisation schemes requiring that power be constantly on will be at a significant disadvantage.

We will see below that, with the island locking at relatively small width, LM stabilisation also has the advantage that the peak power requirement is lower than that for the other envisioned stabilisation schemes. It enables current drive at the island O-point exclusively, where it is most stabilising. In addition to factoring into the time-averaged power requirement, the peak power requirement is an issue for ITER because it will take 3 seconds to switch power between launchers, so that this power must be reserved to the upper EC launcher to allow NTM stabilisation on the required time scale \cite{poli_electron_2018}. The power will therefore not be available for competing demands such as for core heating, sawtooth stabilisation, current profile control, or pumping impurities out of the plasma core. 

A LM stabilisation strategy can be implemented in ITER. The intrinsic error field in ITER will be compensated by external coils, which could be adjusted such that the resulting total error field locks the island O-point in front of the EC wave launcher. Upon detection of a rotating $2/1$ NTM, one could simply wait until it locks before proceeding to stabilise it efficiently. Although not considered here, a combination of RM and LM stabilisation could prove optimal in practice, e.g. to further reduce the island width at locking.

\section{Numerical model of the NTM temporal dynamics}
\label{sec:numerical_model}

For our numerical calculations of the temporal evolution of the NTM island width and rotation frequency, we use the `modified Rutherford equation' (also known as the `generald Rutherford equation') coupled to an equation describing the time evolution of the island rotation.  These equations are perhaps best viewed as scaling models that have been fitted to the available experimental data.  The modified Rutherford equation was originally derived via theoretical calculations of the time evolution of narrow islands in cylindrical geometry. The treatment of the effect of viscosity  on the island rotation has also been simplified \cite{la_haye_effect_2017}. The equations have been modified by incorporating coefficients and other parameters such that the predictions of the equations provide an approximate fit to the available experimental data.  Virtually all of the calculations of ECCD NTM stabilisation for ITER have made use of subsets of these equations \cite{de_lazzari_merits_2009, bertelli_requirements_2011, van_den_brand_integrated_2012, poli_criteria_2015, figini_assessment_2015}. We believe that these equations provide, at present, the most reliable projections of such stabilisation. Nevertheless, there remain uncertainties in these predictions, as discussed further below.

The Modified Rutherford Equation that we use to calculate the time evolution of the island width is
\begin{align}
           0.82 \frac{\tau_r}{r_s} \frac{\mathrm{d}w}{\mathrm{d}t} = & \underbrace{r_s \left( \Delta_0' +{\Delta'_\mathrm{0, wall}(w,\omega)}\right)}_{\text{Classical}} + \underbrace{{2 m \left( \frac{w_\mathrm{vac}}{w} \right)^2 \cos(\phi-\phi_\mathrm{EF})}}_\text{Error field / RMP}    \nonumber \\ 
           & + 2.8\; \frac{j_\mathrm{BS}}{j_\parallel} L_q \left( \underbrace{\frac{2}{3{w}}}_\text{Bootstrap} - \underbrace{\frac{3w_\mathrm{ib}^2}{w^3}}_\text{Polarisation} - \underbrace{\frac{3 \pi^{3/2}}{4 w_\mathrm{dep}} \frac{w_\mathrm{dep}^2}{{w^2}} \eta_\mathrm{NTM} {\eta_\mathrm{stab}}(w,\phi) }_\text{Current drive} \right). 
\label{eq:rutherford_ITER}
\end{align}

 In this study, we focus on the stabilisation of the $2/1$-mode, which is the most challenging to stabilise during the rotating phase. The bootstrap and polarisation coefficients originate from \cite{la_haye_effect_2017}, based on DIII-D's Iter Baseline Scenario (IBS). The wall stabilising term $\Delta_0'(\omega)$ can be found in \cite{nave_mode_1990}, the error field term in \cite{fitzpatrick_interaction_1993}, and the current drive term in \cite{bertelli_requirements_2011}. The classical contribution $\Delta_0'$ is chosen such that the classical term in \eqref{eq:rutherford_ITER} is equal to $r_s \left( \Delta_0' +\Delta'_\mathrm{0, wall}\right) = 1.1$ in the limit of $\omega\ll\tau_w^{-1}$, to obtain the same rotating island evolution as \cite{la_haye_effect_2017}.

In \eqref{eq:rutherford_ITER}, $\tau_r = 273$~s is the local resistive time (where we used $Z_\mathrm{eff} = 1.53$ \cite{polevoi_private_2019}), $r_s = 1.55$~m the minor radius of the rational surface at which the island forms, $m=2$ is the poloidal mode number of the island, $\phi_\mathrm{EF}$ is the phase of the error field term, $j_\mathrm{BS}=72$~kA$/$m$^2$ and $j_\parallel=388$~kA$/$m$^2$ are the local bootstrap and parallel current densities, respectively; $L_q = q/(\mathrm{d}q/\mathrm{d}r) = 0.94$~m is the local shear length, and $w_\mathrm{ib}=0.7$~cm is the ion banana width. In all calculations shown, the stabilisation efficiency $\eta_\mathrm{stab}$ is evaluated exactly using \cite{hegna_stabilization_1997, giruzzi_dynamical_1999, de_lazzari_merits_2009}
\begin{equation}
    \eta_\mathrm{stab} = D_\mathrm{mod} \frac{\int_{-1}^{\infty} \mathrm{d}\Omega \; \langle j_\mathrm{CD} \rangle \frac{\langle \cos(m \xi) \rangle}{\langle 1 \rangle} }{\int_{-1}^{\infty} \mathrm{d}\Omega \; \langle j_\mathrm{CD} \rangle }. \label{eq:eta_stab}
\end{equation}
Here, $\xi$ is a helical angle, $\Omega$ is a label for the island flux surfaces, $\langle\dots\rangle$ is a flux-surface average, and $D_\mathrm{mod}\in [0,1]$ is the modulation parameter for rotating island stabilisation. Values of the stabilisation efficiency range between $-1$ and $1$, corresponding to $\delta$-function current drive profiles at the island separatrix $\Omega = 1$, and at the island O-point $\Omega=-1$, respectively. 
Finally, $\eta_\mathrm{NTM}$ is defined as
\begin{equation}
    \eta_\mathrm{NTM} = \frac{j_\mathrm{CD, max}}{j_\mathrm{BS}} =  \frac{\eta_\mathrm{CD}}{\pi^{3/2} r_s w_\mathrm{dep} j_\mathrm{BS}} P_\mathrm{RF},
\end{equation}
for an EC-driven current density profile that is gaussian with radial width $w_\mathrm{dep}$. We do not consider the impact of radial misalignment of the EC wave with the resonant surface, a previously identified challenge for NTM stabilisation \cite{la_haye_requirements_2008, de_lazzari_merits_2009, urso_asdex_2010, van_den_brand_integrated_2012, poli_criteria_2015, poli_electron_2018}, as it becomes much less severe for the large deposition widths of interest in this study. The current drive efficiency $\eta_\mathrm{CD} = I_\mathrm{CD}/P_\mathrm{RF}$ and unbroadened deposition width are obtained from \cite{bertelli_requirements_2011}. Here, we use the values corresponding to a toroidal launching angle of $20^\circ$ for the lower steering mirror, as it has a narrower deposition width $w_\mathrm{dep}$ than the upper steering mirror; then, $\eta_\mathrm{CD} \approx 6.2$~kA$/$MW and $w_\mathrm{dep0}=2.8$~cm. Broadening is modeled as a multiplicative constant on the deposition width, a broadening factor of $1$ thus corresponding to the unbroadened value. 

The island rotation is evolved in time according to
\begin{equation}
    \frac{\mathrm{d}\omega}{\mathrm{d}t} = \underbrace{\frac{\omega_0 - \omega\left( 1 + C_M \frac{w}{a}\right)}{\tau_{M0}}}_\text{Viscous} - \frac{1}{\tau_{A0}^2} \left( \frac{w}{a} \right)^3 \left[ \underbrace{ \frac{C_1}{m} \frac{\omega \tau_w}{(\omega \tau_w)^2 + 1}}_\text{Resistive wall} + \underbrace{\frac{m^2}{256} \left(\frac{a}{L_q} \right)^2 \left( \frac{w_\mathrm{vac}}{w} \right)^2  \sin(\phi-\phi_\mathrm{EF})}_\text{Error field / RMP} \right], \label{eq:rotation_ITER}
\end{equation}
with the viscous and resistive wall terms from \cite{la_haye_effect_2017}. The error field term, based on \cite{fitzpatrick_interaction_1993}, was added keeping the same island inertia coefficient $\nu=3$ as the resistive wall term. The coefficients $C_M = 12$ and $C_1 = 1/80$ are experimentally fitted to the DIII-D IBS discharge \cite{la_haye_effect_2017} and account for degradation of the momentum confinement and strength of the wall coupling, respectively. Finally, $\tau_w=14$~ms is the wall time corresponding to ITER's blanket modules, $\tau_{M0}=3.7$~s is the original momentum confinement time, and $\tau_{A0}=3\;\mu$s is the Alfv\'en time.

The system of equations \eqref{eq:rutherford_ITER} and \eqref{eq:rotation_ITER} can be solved alongside $\mathrm{d}\phi / \mathrm{d}t = \omega$ to evolve the island width and rotation in time. We will consider an island seeded at a frequency of $0.42$~kHz and a width $w_\mathrm{seed}=3 w_\mathrm{ib} = 2.1$~cm \cite{la_haye_effect_2017}.

Most previous studies of NTM stabilisation for ITER have also made use of the Generalised Rutherford Equation \cite{de_lazzari_merits_2009, bertelli_requirements_2011, van_den_brand_integrated_2012, poli_criteria_2015, figini_assessment_2015}, albeit in a simplified form. More importantly, only one previous study \cite{van_den_brand_integrated_2012} considered the implications of island deceleration and mode locking on rotating island stabilisation, studying in that case the permitted time delay between mode seeding and EC stabilisation.

\begin{figure}[!ht]
    \centering
    \includegraphics[width=1\textwidth]{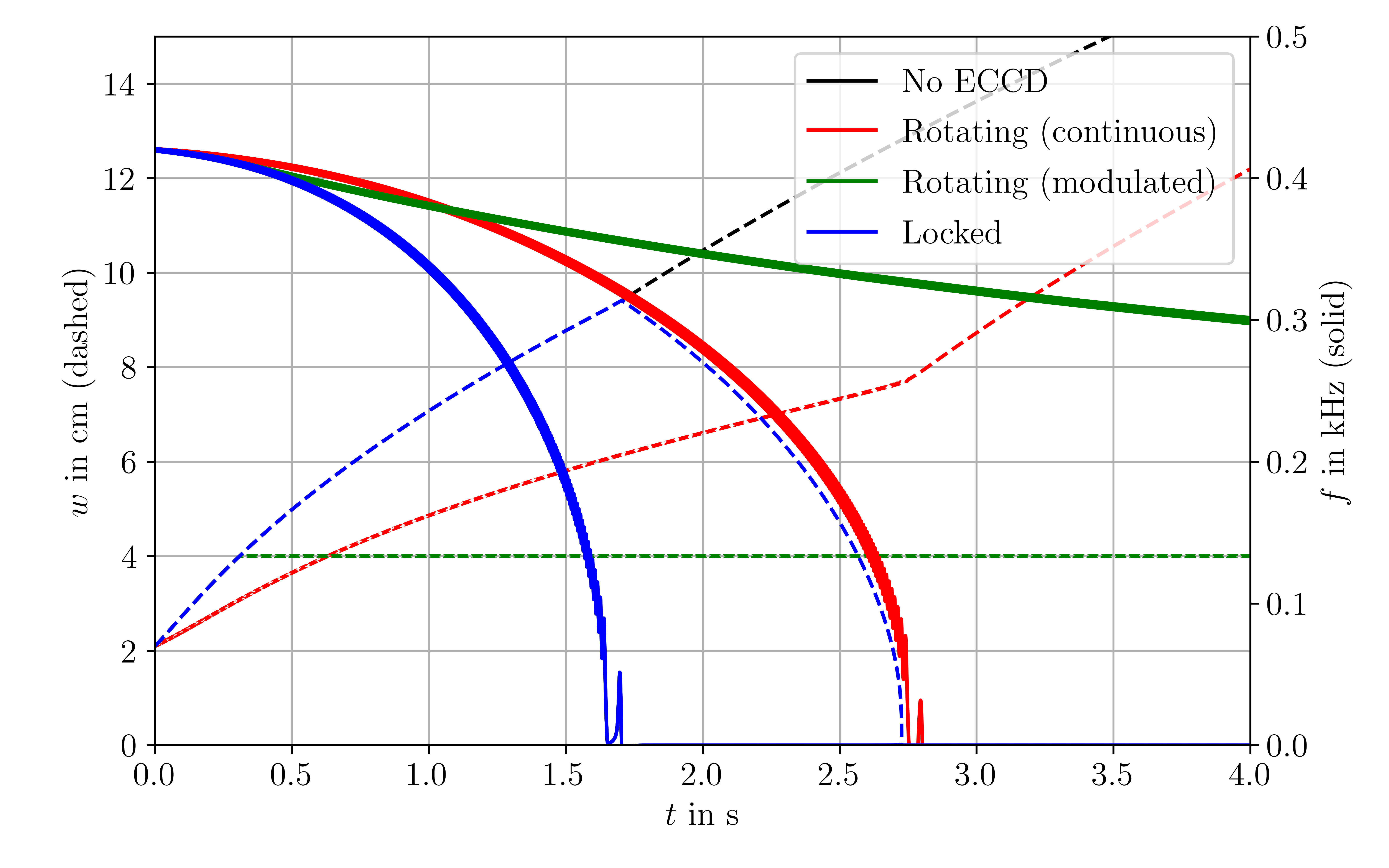}
    \caption{Temporal evolution of NTM island width and rotation frequency for ITER's lower steering mirror, a broadening factor of $3$ and $7.5$~MW of EC power for NTM stabilisation.}
    \label{fig:temporal_evolution_NTM}
\end{figure}

An illustrative temporal evolution of island width and rotation is shown in Fig.~\ref{fig:temporal_evolution_NTM}, assuming a broadening factor of $3$. For RM stabilisation, both preemptive continuous power and modulated power are considered. For preemptive stabilisation, as considered e.g. by \cite{poli_electron_2018}, the power remains turned on at
all times. The modulated power is turned on when the island width exceeds a detection threshold, assumed herein to be $w_\mathrm{detect}=4$~cm, following \cite{poli_electron_2018}. It is assumed that the modulated power is phased perfectly at $50\%$ modulation (which must be maintained while the island is decelerating). LM stabilisation is also considered, for which the power is turned on only after locking of the mode. For the sake of simplicity, a single launcher at $\phi = \phi_\mathrm{EF}$ is assumed.

As can be seen in Fig.~\ref{fig:temporal_evolution_NTM}, in the absence of ECCD stabilisation the island locks after only $\sim 1.7$~s, at a small width of $\sim 9$~cm, corresponding to $\sim 4.5\%$ of the minor radius. With $7.5$~MW employed, RM stabilisation with continuous power does not stabilise the NTM, only delaying the locking by $\sim 1$~s. In contrast, stabilisation of the RM with modulated power and of the LM is successful with $7.5$~MW. For the modulated power case, the island width cannot be reduced below the detection threshold, and the power must remain on to prevent the island from growing and locking. The detection threshold width here is slightly below the critical island width for locking, and the island rotation slows and asymptotes to a lower frequency but does not lock. The power can be turned off for LM stabilisation once the island width is below the instability threshold. Here, the locked island is fully stabilised within only $1$~s after locking.

\section{Peak power requirement for NTM stabilisation}
\label{sec:power}

 We now calculate the peak power requirements for stabilisation using the model for the time evolution of the island discussed in the previous section. For a given EC power, the system of equations \eqref{eq:rutherford_ITER} and \eqref{eq:rotation_ITER} can be evolved in time and it can be assessed whether the given power was sufficient to stabilise the island. Using a bisection method, we evaluate the power requirements for different scenarios and deposition widths, to an accuracy of $0.1$~MW. In ITER, this required EC power must be reserved to the upper launcher because of the 3 seconds that it takes to switch power between launchers [23], and this power is therefore not available for competing demands such as for
core heating, sawtooth stabilisation, stabilisation of the
3/2 NTM, or pumping impurities out of the plasma core. For RM stabilisation with continuous power we focus on preemptive stabilisation, as the peak power requirement is lower compared to the case where the RF is turned on after the island is detected. The calculations for modulated RF turn on the power after $w$ exceeds $4$~cm.  For LMs, two cases are considered with different $t_\mathrm{stab} =2, 10$~s, where $t_\mathrm{stab}$ is the maximum time allowed for stabilisation after locking. While the peak power requirement will be lower for $t_\mathrm{stab} = 10$~s, it may be preferable in practice to opt for faster stabilisation, to reduce the total energy expended to stabilise the mode or to avoid a potential loss of H-mode on a momentum confinement timescale, $\tau_M \sim 2.5$~s at locking in ITER.

\begin{figure}[!ht]
    \centering
    \includegraphics[width=1\textwidth]{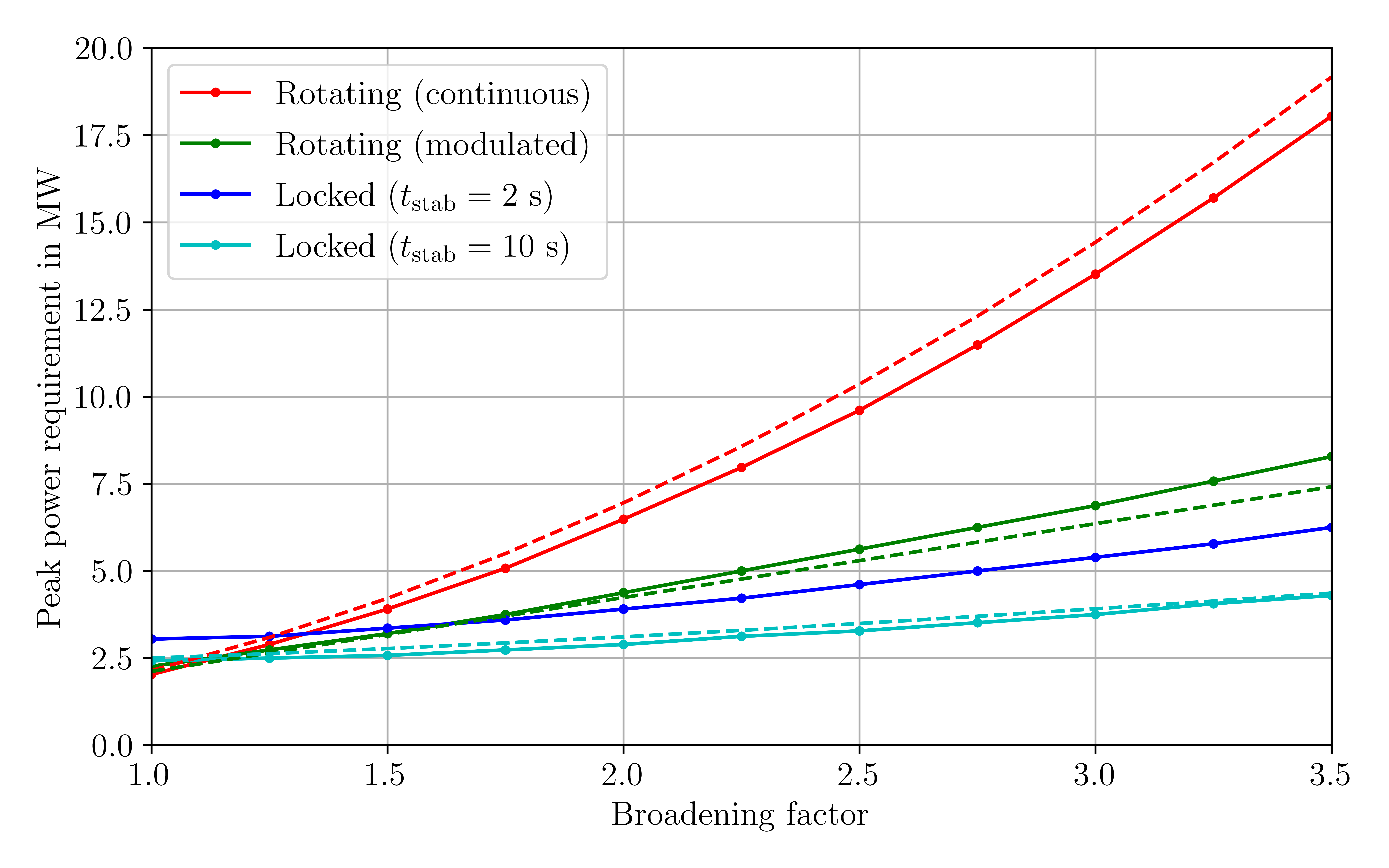}
    \caption{Peak power requirements for different NTM stabilisation scenarios in ITER. The solid lines show the values obtained by evolving the system of equations \eqref{eq:rutherford_ITER} and \eqref{eq:rotation_ITER} in time. The dashed lines show the predictions of approximate formulae derived in Section \ref{sec:island_stab_large_wdep}.}
    \label{fig:plot_broadening_LSM}
\end{figure}

The power requirements for different values of the broadening factor are shown in Fig.~\ref{fig:plot_broadening_LSM}. Even without broadening, LM stabilisation requires a similar amount of power to RM stabilisation. With the current driven predominantly near the island O-point for the LM, the increased stabilisation efficiency $\eta_\mathrm{aux}$ counterbalances the $1/w^2$ dependence of the CD stabilisation term in (\ref{eq:rutherford_ITER}). As the broadening is increased, the power requirement increases rapidly for RM stabilisation with continuous power, somewhat slower for modulated power and slowest for LM stabilisation. At the broadening factor of $\sim 3$ predicted for ITER \cite{snicker_effect_2018}, Fig.~\ref{fig:plot_broadening_LSM} shows that NTM stabilisation with the LSM would require only $3-5$~MW for a LM, while it would require $7-13$~MW for a RM.

\begin{figure}[!ht]
    \centering
    \includegraphics[width=1\textwidth]{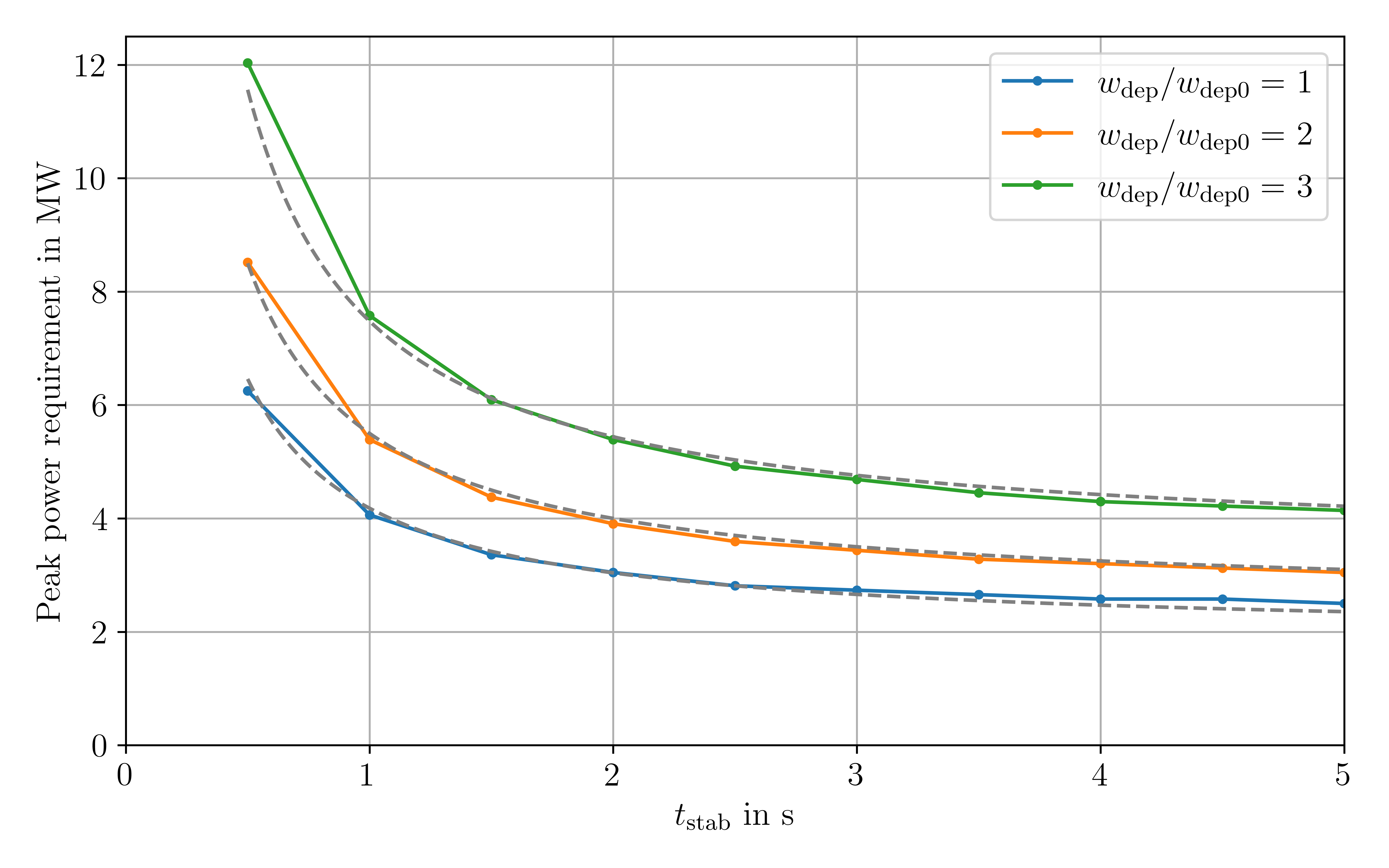}
    \caption{Peak power requirement for locked mode stabilisation in ITER, as a function of the broadening factor and $t_\mathrm{stab}$, the time required to stabilise the island after locking. The grey dashed lines correspond to a $P_\mathrm{RF} \propto 1 + (t_\mathrm{stab}/1.2 \mathrm{ s})^{-1}$ dependence.}
    \label{fig:plot_tstab_LSM}
\end{figure}

For locked mode stabilisation, the power requirements are shown in Fig.~\ref{fig:plot_tstab_LSM} as a function of $t_\mathrm{stab}$, for various broadening values. The power requirement is shown to increase approximately as $P_\mathrm{RF} \propto 1 + (t_\mathrm{stab}/1.2 \mathrm{ s})^{-1}$, independent of the broadening factor. The increase in the peak power requirement to stabilise the locked island within a momentum confinement in ITER thus remains moderate. As the energy required to stabilise the island is $P_\mathrm{RF} t_\mathrm{stab}$, the aforementioned scaling suggests that the island should be stabilised as quickly as possible after locking to reduce the energy expendited to stabilise the mode.

To summarise, for projected EC deposition profile widths, power requirements for RM stabilisation via continuous RF are predicted to be prohibitively large.  Also, stabilisation via either modulated RF or preemptive continuous RF require the power to remain on throughout the shot, with corresponding impact on the required time averaged power and therefore on $Q$. LM stabilisation does not require the RF to remain on, and the peak power requirements are also lower.

\section{Perspective on the power requirements from approximate analytical solutions}
\label{sec:island_stab_large_wdep}

In Section \ref{sec:power}, numerical time dependent solutions of the modified Rutherford equation assuming ITER Scenario 2 equilibrium parameters at the $q=2$ surface were used to determine the power requirements associated with several alternative island stabilisation strategies. The Scenario 2 equilibrium is the reference equilibrium that is considered the leading candidate for achieving the highest value of $Q$ in ITER, and the associated parameters have been used in the calculations that have driven the design of the ITER EC launchers \cite{ramponi_iter_2007,henderson_overview_2008,figini_assessment_2015,poli_criteria_2015}. In this section we use approximate analytical solutions to obtain a comparison of the power requirements for a broader set of parameters. Relatively simple analytical expressions for the peak power required to stabilise the NTM are obtained. The dashed lines in Fig.~\ref{fig:plot_broadening_LSM} show the predictions of these expressions when evaluated for the parameters used in Section \ref{sec:power}. It can be seen that the expressions provide an adequate approximation to the numerical results.



To obtain a broader picture of the relative power requirements, it is instructive to consider the general form of the modified Rutherford equation:
\begin{equation}
    \frac{\mathrm{d}w}{\mathrm{d}t} = c_\mathrm{cl} + \frac{c_\mathrm{BS}}{w} + \frac{c_\mathrm{EF}}{w^2} - \frac{c_\mathrm{pol}}{w^3} - c_\mathrm{RF} P_\mathrm{RF} \frac{\eta_\mathrm{stab}(w)}{w^2}, \label{eq:GRE_general_coeffs}
\end{equation}
where $c_\mathrm{cl}$  corresponds to the classical contribution (including conducting wall stabilisation \cite{nave_mode_1990}), $c_\mathrm{BS}$ to the bootstrap drive leading to growth of the NTM, $c_\mathrm{EF}$ to driven reconnection from the error field \cite{fitzpatrick_interaction_1993} relevant only to the LM case, $c_\mathrm{pol}$ to the polarisation current effect stabilising the island at small widths. The last term with $c_\mathrm{RF}$ represents the contribution from RF current drive, which is linearly proportional to the wave power $P_\mathrm{RF}$ and the stabilisation efficiency $\eta_\mathrm{stab}$, introduced in Section \ref{sec:numerical_model}. The values of the coefficients in the ITER case previously considered can be obtained by comparing with Eq.~\ref{eq:rutherford_ITER}. An analytic treatment of the power requirements will require analytic approximations to $\eta_\mathrm{stab}$ that are appropriate for each of the three stabilisation strategies that we consider.

Considering first the stability properties of the island in the absence of RF and in the absence of the error field term, it can be seen from Eq.~\ref{eq:GRE_general_coeffs} that the island is metastable.  That is, it is stable when the island width is small but may become unstable at larger values of the island width.  For an unstable NTM, we define the marginal island width, $w_\mathrm{marg}$, to be the threshold island width at which the island becomes unstable in the absence of the RF and error field terms, $w_\mathrm{marg} \approx \sqrt{c_\mathrm{pol}/c_\mathrm{BS}}$. Here we have neglected the classical term, whose relative contribution is generally relatively small for the narrow islands under consideration. Eq.~\ref{eq:rutherford_ITER} gives $w_\mathrm{marg} \approx 3w_\mathrm{ib}/\sqrt{2}$.

If the ECCD term in Eq.~\ref{eq:GRE_general_coeffs} is sufficiently large to drive the island width below $w_\mathrm{marg}$, the ECCD at the rational surface can be turned off until the next NTM triggering event. If the ECCD power is sufficient to stabilise the island, but not sufficient to reduce its width below $w_\mathrm{marg}$, the ECCD must be maintained at the rational surface throughout the shot, with potential implications for $Q$. For the ITER reference design we estimated $w_\mathrm{ib} \approx 0.7$ cm, giving $w_\mathrm{marg} \approx 1.5$ cm. Without the EC power and error field terms, the time derivative of the island width peaks at $w_\mathrm{peak} \approx \sqrt{3} w_\mathrm{marg}$. For $w_\mathrm{ib} \approx 0.7$ cm, this gives $w_\mathrm{peak} \approx 2.6$ cm. By comparison, the ECCD deposition width in the absence of broadening is calculated to be 2.8 cm, and is 8.4 cm for a broadening factor of 3. 

\subsection{Rotating island stabilisation}

Let us first consider island stabilisation during the rotating phase. In this case, we may neglect the error field contribution to the Rutherford equation \eqref{eq:GRE_general_coeffs}. For rotating island stabilisation with continuous power, we consider a preemptive stabilisation scenario, where the RF power is turned on at all times and thus acts to stabilise the island immediately upon seeding. In the case of modulated power, we instead consider stabilisation immediately upon detection of the island. These choices may be regarded as best-case scenarios for rotating island stabilisation, as requiring stabilisation at larger widths than considered here can only increase the power requirement for stabilisation. As the width at seeding and the width at detection will generally be smaller than the deposition width when broadening effects are included, the above considerations motivate the use of a set of approximations for $\eta_\mathrm{stab}$ that are valid for $w \lesssim w_\mathrm{dep}$. 

\begin{figure}[!ht]
    \centering
    \includegraphics[width=1\textwidth]{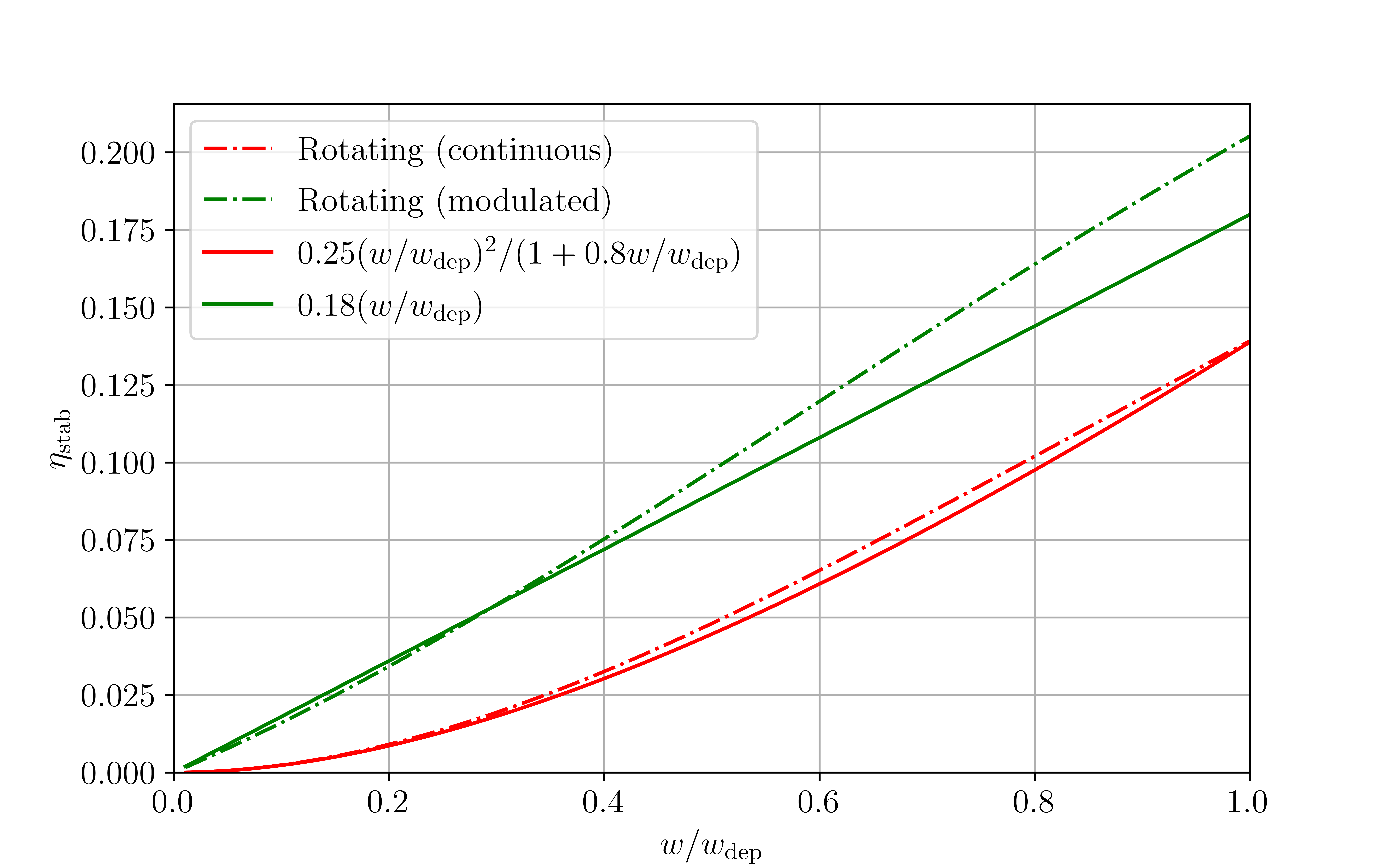}
    \caption{Current drive stabilisation efficiency for rotating island stabilisation, and comparison to approximate formulae.}
    \label{fig:plot_check_eta_CD_rot}
\end{figure}

The dashed lines in Fig.~\ref{fig:plot_check_eta_CD_rot} show the stabilisation efficiency values for $w_\mathrm{dep} \ge w$ evaluated directly from Eq.~\eqref{eq:eta_stab}, with the different scenarios corresponding to different angular dependencies of $\eta_\mathrm{stab}$. The stabilisation efficiency can be approximated as $\eta_\mathrm{stab}^\mathrm{RM, ctd.} \approx 0.25 (w/w_\mathrm{dep})^2 / ( 1 + 0.8 w/w_\mathrm{dep} )$ for RM stabilisation with continuous power, and as $\eta_\mathrm{stab}^\mathrm{RM, mod.} \approx 0.18 (w/w_\mathrm{dep})$ for RM stabilisation with $50\%$ modulated power, and the corresponding curves are shown as solid lines in Fig.~\ref{fig:plot_check_eta_CD_rot}. The relative errors between the approximate formulae and exact evaluation of $\eta_\mathrm{stab}$ remain below $15\%$ in the range $w/w_\mathrm{dep}\in[0,1]$.

\subsubsection{Rotating island stabilisation with continuous power}

~\newline

For rotating island stabilisation with continuous power, we consider preemptive stabilisation, i.e. the power is turned on at all times, and acts to stabilise the mode immediately after seeding.

In the case of RM stabilisation with continuous power, we can approximate $\eta_\mathrm{stab}$ as $\eta_\mathrm{stab} = c_\mathrm{cts} w^2 / (1 + w/w_\mathrm{cts})$. As $w_\mathrm{cts} \rightarrow \infty$, the wave-driven current contribution to the GRE becomes a constant, and the island growth peaks at the width $w_\mathrm{peak}$ of the case without RF stabilisation. For finite $w_\mathrm{cts}$, the peak growth is shifted to smaller $w <w_\mathrm{peak}$, but this shift is small in practice because of the strong $w^{-3}$-dependence of the polarisation current contribution. For the GRE coefficients employed in \eqref{eq:rutherford_ITER}, the shift being small only requires $w_\mathrm{marg}^2 \ll r_s w_\mathrm{dep} j_\mathrm{EC}/j_\mathrm{BS}$, which is generally satisfied. Then, because $\eta_\mathrm{stab}/w^2$ is a monotonically decreasing function of $w$ in this case, the criterion for reducing the island width below $w_\mathrm{marg}$ can simply be obtained by requiring
\begin{equation}
    \frac{\mathrm{d}w}{\mathrm{d}t}\bigg|_{w = \mathrm{min}(w_\mathrm{seed}, w_\mathrm{peak})} < 0,
\end{equation}
or equivalently
\begin{equation}
    P_\mathrm{RF}^\mathrm{cts} > \frac{1}{c_\mathrm{cts} c_\mathrm{RF}} \left(1+ \frac{w}{w_\mathrm{cts}} \right) \left[ c_\mathrm{cl} + \frac{c_\mathrm{BS}}{w} \left( 1 - \frac{w_\mathrm{marg}^2}{w^2} \right) \right] \bigg|_{w = \mathrm{min}(w_\mathrm{seed}, w_\mathrm{peak})}.  \label{eq:criterion_c_CD_RM_ctd}
\end{equation}
The power requirement is thus somewhat relaxed when the seed island width is below the peak island width, as is the case in our ITER simulations, where $w_\mathrm{seed} \approx 2.1$~cm~$< w_\mathrm{peak} \approx 2.6$~cm.

Associating the coefficients in \eqref{eq:GRE_general_coeffs} with the terms in \eqref{eq:rutherford_ITER}, the criterion \eqref{eq:criterion_c_CD_RM_ctd} for the case $w_\mathrm{seed} < w_\mathrm{peak}$ leads to
\begin{align}
    \nonumber P_\mathrm{RF}^\mathrm{cts} & > \frac{4}{3\cdot0.25} \frac{j_\mathrm{BS}w_\mathrm{dep}^2}{\eta_\mathrm{CD}} \left(1 + 0.8\frac{ w_\mathrm{seed}}{w_\mathrm{dep}} \right) \left[ \frac{j_\parallel}{2.8 j_\mathrm{BS}}\frac{r_s}{L_q} r_s \left( \Delta_0' + \Delta_\mathrm{0,wall}'\right) + \frac{2}{3} \frac{r_s}{w_\mathrm{seed}} \left( 1 - \frac{w_\mathrm{marg}^2}{w_\mathrm{seed}^2} \right) \right] \\
    & \approx  1.33 \left(1 + \frac{1.68\;\mathrm{cm}}{w_\mathrm{dep}} \right) \left(\frac{w_\mathrm{dep}}{2.77\;\mathrm{cm}} \right)^2 \;\mathrm{MW} \label{eq:criterion_P_EC_cts_ITER}
\end{align}
Here, we assumed a small destabilising contribution from the classical term, $r_s \left( \Delta_0' + \Delta_\mathrm{0,wall}'\right) = 1.1$, following \cite{la_haye_effect_2017}, which increases the power requirement by $\approx 14\%$. The power requirement increases quadratically with the deposition width for large $w_\mathrm{dep}$, a less favourable scaling than the other stabilisation strategies, as shown below.

Note that the criterion \eqref{eq:criterion_c_CD_RM_ctd} depends on the details of the small island width stabilisation by the polarisation current, as the approximately constant current drive contribution in \eqref{eq:GRE_general_coeffs} cannot compete with the $1/w$-dependent bootstrap drive at very small island widths. The situation is markedly different in the cases of RM stabilisation with modulated power and of LM stabilisation.

\subsubsection{Rotating island stabilisation with modulated power}

~\newline

In this case, the stabilisation efficiency may be approximated as $\eta_\mathrm{stab} = c_\mathrm{mod} w$, and the current drive contribution can compete with the bootstrap term at small island widths. Generally, the classical contribution to the Rutherford equation is expected to be small relative to other terms for the relatively small island widths of relevance here, and it is neglected here.

As discussed in Sections 4 and 5, the island width must exceed the detection threshold $w_\mathrm{detect}$ to enable synchronisation of the modulated power. Due to the existence of the detection threshold and the need for synchronisation, islands can be reduced in size but cannot be fully stabilised with modulated power \cite{hender_chapter_2007}. Furthermore, if the threshold for accurately determining the island phase were to exceed the critical island width $w_\mathrm{crit} \approx 4.5$~cm in ITER \cite{la_haye_effect_2017} above which the island is on course for locking, modulation would only delay the locking event. Even for $w_\mathrm{detect} < w_\mathrm{crit}$, practical issues might arise due to the small amount of time available to aim and turn on the EC power before locking occurs, as investigated in \cite{van_den_brand_integrated_2012}. Here, we turn on the EC power immediately after the detection threshold is exceeded, and assume perfectly synchronised $50\%$ modulation. The stabilisation is considered successful in this case if the island remains at $w_\mathrm{detect}$.

In the regime $w_\mathrm{detect} \lesssim w_\mathrm{dep}$, stabilisation therefore requires
\begin{equation}
    P_\mathrm{RF}^\mathrm{mod} > \frac{c_\mathrm{BS}}{c_\mathrm{mod} c_\mathrm{RF} }\left( 1 - \frac{w_\mathrm{marg}^2}{w_\mathrm{detect}^2} \right). \label{eq:criterion_c_CD_RM_mod}
\end{equation}
The power requirement for stabilisation via modulated RF can be significantly reduced if  $w_\mathrm{detect} \sim w_\mathrm{marg}$. Such a low threshold for detecting the island, and for accurately determining its phase in real time, may however be difficult to realise in practice \cite{poli_electron_2018}.

For rotating island stabilisation with modulated power, again associating the coefficients in \eqref{eq:GRE_general_coeffs} with the terms in \eqref{eq:rutherford_ITER}, Eq.~\eqref{eq:criterion_c_CD_RM_mod} yields
\begin{equation}
    P_\mathrm{RF}^\mathrm{mod} > \left(\frac{j_\mathrm{BS} r_s w_\mathrm{dep}}{\eta_\mathrm{CD}}\right) \frac{8}{9\cdot 0.18} \left[1 - \left(\frac{ w_\mathrm{marg}}{w_\mathrm{detect}}\right)^2 \right] \approx 2.6 \left[1 - \left(\frac{ 1.5\;\mathrm{cm}}{w_\mathrm{detect}}\right)^2 \right] \left(\frac{w_\mathrm{dep}}{2.77\;\mathrm{cm}} \right) \;\mathrm{MW} \label{eq:criterion_P_EC_mod_ITER}
\end{equation}
The linear scaling with the deposition width makes modulation the method of choice for rotating island stabilisation when the broadening is large. Note that the previously assumed detection threshold $w_\mathrm{detect}=4~$cm leads to a reduction in the power requirement by $\approx 14\%$.

\subsection{Locked island stabilisation}

To obtain the power requirement for locked island stabilisation, we require an approximation for $\eta_\mathrm{stab}$ that is valid over a broader range. If the error field is large, the power requirement is determined by requiring the island be stabilised at small widths. If the error field is small, the power requirement is instead set by requiring the island be stabilised at the width at locking. The width at locking may be large relative to the deposition width, even when broadening effects are included.

\begin{figure}[!ht]
    \centering
    \includegraphics[width=1\textwidth]{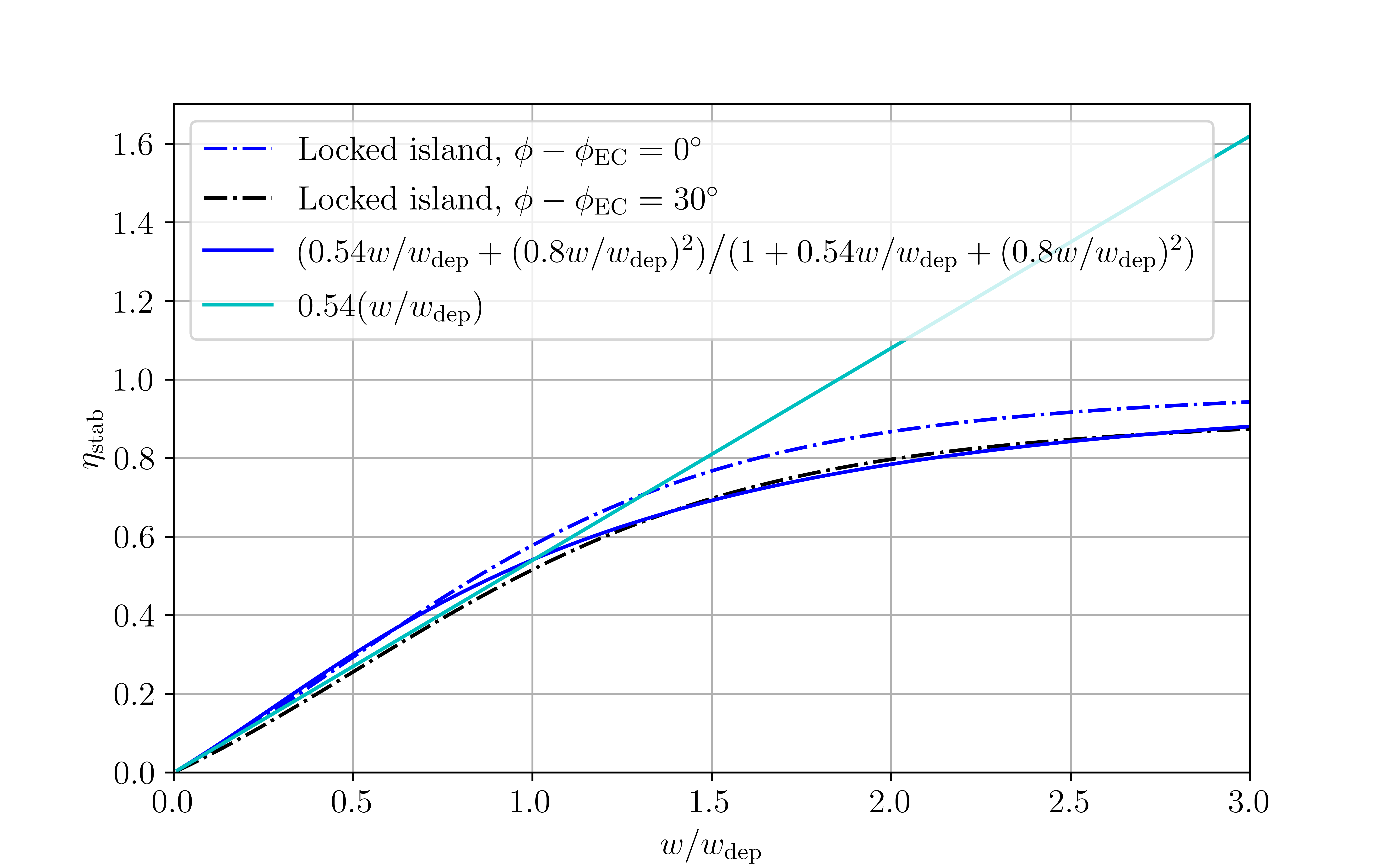}
    \caption{Current drive stabilisation efficiency for locked island stabilisation, and comparison to approximate formulae.}
    \label{fig:plot_check_eta_CD_lok}
\end{figure}

As shown in Fig.~\ref{fig:plot_check_eta_CD_lok}, the stabilisation efficiency may be approximated as 
\begin{equation}
    \eta_\mathrm{stab}^\mathrm{LM} \approx \frac{0.54 w/w_\mathrm{dep} + \left(0.8 w/w_\mathrm{dep}\right)^2}{1 + 0.54 w/w_\mathrm{dep} + \left(0.8 w/w_\mathrm{dep}\right)^2}, \label{eq:approximate_eta_stab_LM}
\end{equation}
which can be simplified to $\eta_\mathrm{stab}^\mathrm{LM} \approx 0.54 (w/w_\mathrm{dep})$ in the large deposition width limit $w \lesssim w_\mathrm{dep}$, and also correctly asymptotes to $\eta_\mathrm{stab}^\mathrm{LM} \approx 1$ for $w \gg w_\mathrm{dep}$. The relative error between the approximate formula and exact evaluation of $\eta_\mathrm{stab}$ remains below $11\%$ in the range $w/w_\mathrm{dep}\in[0,\infty]$. Finally, note that the stabilisation efficiency remains high even when the island is imperfectly aligned with the EC-driven wave in the angular direction, as shown in Fig.~\ref{fig:plot_check_eta_CD_lok} for $\phi-\phi_\mathrm{EC}=30^\circ$.

Let us first consider the regime where the power requirement is set at small island widths, such that the large broadening of the stabilisation efficiency may be considered, $\eta_\mathrm{stab}^\mathrm{LM} = c_\mathrm{LM} w$. Then, again assuming the classical $\Delta'$ contribution $c_\mathrm{cl}$ is negligible, the island growth in time \eqref{eq:GRE_general_coeffs} is given by
\begin{equation}
        \frac{\mathrm{d}w}{\mathrm{d}t} = \frac{c_\mathrm{BS} - c_\mathrm{RF}P_\mathrm{RF} c_\mathrm{LM}}{w} + \frac{c_\mathrm{EF}}{w^2} - \frac{c_\mathrm{pol}}{w^3}.
\end{equation}
Requiring the island to be stabilised for all island widths then leads to the stabilisation criterion
\begin{equation}
    P_\mathrm{RF}^\mathrm{LM} > \frac{1}{c_\mathrm{RF}c_\mathrm{LM}} \left( c_\mathrm{BS} + \frac{c_\mathrm{EF}^2}{4 c_\mathrm{pol}} \right). \label{eq:criterion_P_RF_LM_large_wdep}
\end{equation}
Using the coefficients in Eq.~\ref{eq:rutherford_ITER}, Eqs. \ref{eq:criterion_P_RF_LM_large_wdep} and \ref{eq:criterion_error field_GRE} leads to
\begin{equation}
    P_\mathrm{RF}^\mathrm{LM} > \left(\frac{j_\mathrm{BS} r_s w_\mathrm{dep}}{\eta_\mathrm{CD}}\right) \frac{8}{9\cdot 0.54} \left[ 1 + \left( \frac{w_\mathrm{vac}}{w_\mathrm{vac}^\mathrm{power}} \right)^4 \right] \approx 0.87 \left(\frac{w_\mathrm{dep}}{2.77\;\mathrm{cm}} \right) \left[ 1 + \left( \frac{w_\mathrm{vac}}{5\;\mathrm{cm}} \right)^4 \right] \;\mathrm{MW}, \label{eq:criterion_P_EC_LM_ITER_large_wdep},
\end{equation}
where we used
\begin{equation}
    w_\mathrm{vac}^\mathrm{power} \equiv \sqrt{\frac{2.8}{\sqrt{m}} \frac{j_\mathrm{BS}}{j_\parallel} w_\mathrm{ib} L_q } \approx 5\;\mathrm{cm}. \label{eq:wvac_power}
\end{equation}
Here, we assumed the island to be locked at the error field location $\phi=\phi_\mathrm{EF}$. The power requirement for locked island scales linearly with the deposition width, just like rotating island stabilisation with modulated power. However, for small enough vacuum island widths, the power requirement for locked island stabilisation is lower by a factor $\approx 3$ owing to the higher stabilisation efficiency, compare Figs.~\ref{fig:plot_check_eta_CD_rot}~and~\ref{fig:plot_check_eta_CD_lok}.

When equality holds in \eqref{eq:criterion_P_RF_LM_large_wdep}, $\mathrm{d}w/\mathrm{d}t = 0$ at $w = w_\mathrm{peak} = 2 c_\mathrm{pol}/c_\mathrm{EF}$, which may be large for small error fields. The derived stabilisation criterion may then break, as the simplified linear scaling of the stabilisation efficiency employed here is a gross overestimation in the region $w \gtrsim 1.5 w_\mathrm{dep}$, as the cyan curve in Fig.~\ref{fig:plot_check_eta_CD_lok} demonstrates. 

For values of $w_\mathrm{peak}$ exceeding the width at locking $w_\mathrm{lock}$, we need only require the island be stabilised at $w_\mathrm{lock}$, as the power is assumed to be turned on immediately after locking. We now allow for narrower deposition widths $ w_\mathrm{dep} < w_\mathrm{lock}$, and retain the classical contribution $c_\mathrm{cl}$, as its size relative to the bootstrap term grows more important for larger island widths. Requiring the island to be stabilised for $w=w_\mathrm{lock}$ then yields the stabilisation criterion
\begin{equation}
    P_\mathrm{RF}^\mathrm{LM} > \frac{w_\mathrm{lock}^2}{c_\mathrm{RF}\; \eta_\mathrm{stab}^{LM}(w=w_\mathrm{lock})} \left( c_\mathrm{cl} + \frac{c_\mathrm{BS}}{w_\mathrm{lock}} + \frac{c_\mathrm{EF}}{w_\mathrm{lock}^2} \right)\label{eq:criterion_P_RF_LM_general_wdep}
\end{equation}
where $\eta_\mathrm{stab}^{LM}(w=w_\mathrm{lock})$ may be approximated using \eqref{eq:approximate_eta_stab_LM}. Note the polarisation current contribution is neglected in \eqref{eq:criterion_P_RF_LM_general_wdep}, assuming $w_\mathrm{lock}\gg w_\mathrm{marg}$. Using the ITER parameters from \eqref{eq:rutherford_ITER}, the power requirement \eqref{eq:criterion_P_RF_LM_general_wdep} may be expressed as
\begin{align}
    \nonumber P_\mathrm{RF}^\mathrm{LM} & > \left(\frac{j_\mathrm{BS} r_s w_\mathrm{dep}}{\eta_\mathrm{CD}}\right) \frac{8}{9} \frac{w_\mathrm{lock}}{w_\mathrm{dep}} \frac{1}{\eta_\mathrm{stab}^{LM}(w=w_\mathrm{lock})} \left[ 1 + 0.54 \frac{w_\mathrm{lock}}{L_q} \frac{j_\parallel}{j_\mathrm{BS}} \left( r_s \Delta_0' + 2m \left(\frac{w_\mathrm{vac}}{w_\mathrm{lock}}\right)^2 \right) \right] \\
    & \approx \frac{2.1\; \mathrm{MW}}{\eta_\mathrm{stab}^{LM}(w=w_\mathrm{lock})} \left[ 1 + 0.23 \left( \frac{w_\mathrm{vac}}{5\;\mathrm{cm}} \right)^2 \right] \label{eq:criterion_P_EC_LM_ITER_general_wdep},
\end{align}
where $\eta_\mathrm{stab}^{LM}(w=w_\mathrm{lock})$ may be approximated using \eqref{eq:approximate_eta_stab_LM}, and introduces a dependence on $w_\mathrm{dep}$ for $w_\mathrm{lock} \lesssim w_\mathrm{dep}$.

\subsection{Comparison of the power requirements for the various stabilisation strategies}
\label{sec:ITER_power_requirements}

Comparing the peak power requirements in Fig.~\ref{fig:plot_broadening_LSM}, it can be seen that locked mode stabilisation fares best at large broadening factor values, as expected. Even when the mode is required to be stabilised quickly within $2$~s after locking, the power requirement for a broadening factor of $3$ remains modest at $P \approx 5$~MW. At the same broadening factor value, the rotating island stabilisation schemes require $\approx 7$~MW with $50\%$ modulation, and $\approx 13$~MW without. Furthermore, the calculated power requirements are seen to agree well with the derived criteria \eqref{eq:criterion_P_EC_cts_ITER}, \eqref{eq:criterion_P_EC_mod_ITER}, and \eqref{eq:criterion_P_EC_LM_ITER_general_wdep}.

As discussed above, rotating island stabilisation with modulated power cannot fully stabilise the island, such that the EC power cannot be turned off, and the average power $\langle P_\mathrm{RF}^\mathrm{mod} \rangle = P_\mathrm{RF}^\mathrm{mod} D_\mathrm{mod}$. Similarly, when preemptive stabilisation is considered for rotating island stabilisation with continuous power, the EC power remains on at all times, such that $\langle P_\mathrm{RF}^\mathrm{cts} \rangle = P_\mathrm{RF}^\mathrm{cts}$.

For locked island stabilisation, the island can be fully stabilised, after which the power can be turned off until the next seeding and locking event. For locked island stabilisation to be successful, one should require both derived criteria, \eqref{eq:criterion_P_RF_LM_large_wdep} and \eqref{eq:criterion_P_RF_LM_general_wdep}, to be satisfied. One might further desire fast stabilisation of the mode, which could avert loss of the background plasma rotation and thus preserve H-mode. For the ITER case, the increase in the peak power requirement to stabilise the LM within a time $t_\mathrm{stab}$ was shown in Fig.~\ref{fig:plot_tstab_LSM} to scale as 
\begin{equation}
    P_\mathrm{RF}(t_\mathrm{stab}) \approx P_\mathrm{RF}(\infty) \left( 1 + t_0 / t_\mathrm{stab} \right), \label{eq:power_requirement_inst_finite_time}
\end{equation}
where the constant $t_0 \approx 1.2$~s in the ITER case. In \eqref{eq:power_requirement_inst_finite_time}, $P_\mathrm{RF}(\infty)$ may be obtained from the previously derived criteria \eqref{eq:criterion_P_RF_LM_large_wdep} and \eqref{eq:criterion_P_RF_LM_general_wdep}. For the choice of vacuum island width $w_\mathrm{vac} = 1.25$~cm and the broadening factors considered for the ITER case in Fig.~\ref{fig:plot_broadening_LSM}, the power requirement is set solely by \eqref{eq:criterion_P_RF_LM_general_wdep}. Using \eqref{eq:power_requirement_inst_finite_time}, assuming a time $t_\mathrm{seed}$ between seeding events and a time $t_\mathrm{stab}$ for stabilisation, the average power requirement is obtained as
\begin{equation}
    \langle P_\mathrm{RF} \rangle = \frac{P_\mathrm{RF}(t_\mathrm{stab}) t_\mathrm{stab}}{t_\mathrm{stab} + t_\mathrm{seed} + t_\mathrm{lock}} \approx P_\mathrm{RF}(\infty) \frac{t_\mathrm{stab}+t_0}{t_\mathrm{stab}+t_\mathrm{lock}+t_\mathrm{seed}}. \label{eq:power_requirement_average_finite_time}
\end{equation}
When $t_0 < t_\mathrm{lock}+t_\mathrm{seed}$, as is the case in ITER, it is advisable to stabilise the island as quickly as possible after locking. In practice, there will however be limitations on the amount of peak power available. Generally, one will thus want to stabilise the island quickly enough so as to reduce the averaged power requirement \eqref{eq:power_requirement_average_finite_time} and also to enable a fast return to the high-performance island-free state, while keeping the peak power \eqref{eq:power_requirement_inst_finite_time} within the allowed limit. Sawteeth are predicted to have periods of several tens of seconds in ITER \cite{hender_chapter_2007}. If sawteeth are the main cause of NTM seeding in ITER, a locked island stabilisation strategy could permit a substantial reduction of the averaged EC power requirement for NTM stabilisation, and have a corresponding impact on ITER's fusion gain.

\subsection{Error field requirements for locked mode stabilisation}
\label{sec:ITER_error_field_requirements}

For LM stabilisation to be effective, the error field must also be small enough so as not to dominate the criterion \eqref{eq:criterion_P_RF_LM_large_wdep}, i.e.
\begin{equation}
    c_\mathrm{EF} \lesssim \sqrt{2 c_\mathrm{BS} c_\mathrm{pol}} \label{eq:criterion_error field_GRE}.
\end{equation}
For the ITER parameters of \eqref{eq:rutherford_ITER}, this criterion may be expressed as
\begin{equation}
    w_\mathrm{vac} \lesssim w_\mathrm{vac}^\mathrm{power} \approx 5\;\mathrm{cm}, \label{eq:criterion_w_vac_stab}
\end{equation}
where $w_\mathrm{vac}^\mathrm{power}$ was defined in \eqref{eq:wvac_power}.

The error field should also be large enough to effectively trap the island at the desired phase and enable its stabilisation. For a steady-state solution with vanishing rotation frequency, the phase is determined by a balance between the viscous torque $T_\mathrm{visc}$ and error field torque $T_\mathrm{EF}$. Then, the error field should be large enough so as to satisfy
\begin{equation}
    T_\mathrm{EF} \gtrsim T_\mathrm{visc} \label{eq:criterion_T_EF_locking}
\end{equation}
e.g. for $|\phi - \phi_\mathrm{EF}| \approx 30^\circ$, as it was shown in Fig.~\ref{fig:plot_check_eta_CD_lok} that the stabilisation efficiency remains high for such phase misalignments. This criterion should be satisfied for all island widths between the width at locking $w_\mathrm{lock}$ and the marginal island width $w_\mathrm{marg}$ below which small island width effects stabilise the island, without the need for current drive stabilisation. The existence of a range of error field values for which \eqref{eq:criterion_error field_GRE} and \eqref{eq:criterion_T_EF_locking} are simultaneously satisfied is a necessary condition for the practicability of LM stabilisation. 

The criterion \eqref{eq:criterion_T_EF_locking} may also be expressed as a criterion on the vacuum island width. Using \eqref{eq:rotation_ITER} with $\phi-\phi_\mathrm{EF}=30^\circ$ and $\omega=0$,
\begin{equation}
    w_\mathrm{vac} > w_\mathrm{vac}^\mathrm{place}(w) \equiv \frac{16}{m} w \frac{L_q}{a} \left( \frac{a}{w} \right)^{3/2} \sqrt{\frac{2\omega_0 \tau_{A0}^2}{\tau_{M0}}}. \label{eq:criterion_w_vac_lock}
\end{equation}
To ensure the island initially locks at the desired phase, the vacuum island width must satisfy $w_\mathrm{vac} > w_\mathrm{vac}^\mathrm{place}(w_\mathrm{lock}) \approx 0.4$~cm, as $w_\mathrm{lock} \approx 9$~cm. To ensure the island remains locked at the desired phase until the marginal island width is reached, the vacuum island width must satisfy $w_\mathrm{vac} > w_\mathrm{vac}^\mathrm{place}(w_\mathrm{marg}) \approx 1$~cm, as $w_\mathrm{marg} = 3 w_\mathrm{ib}/\sqrt{2} \approx 1.5$~cm.

\begin{figure}[!ht]
    \centering
    \includegraphics[width=1\textwidth]{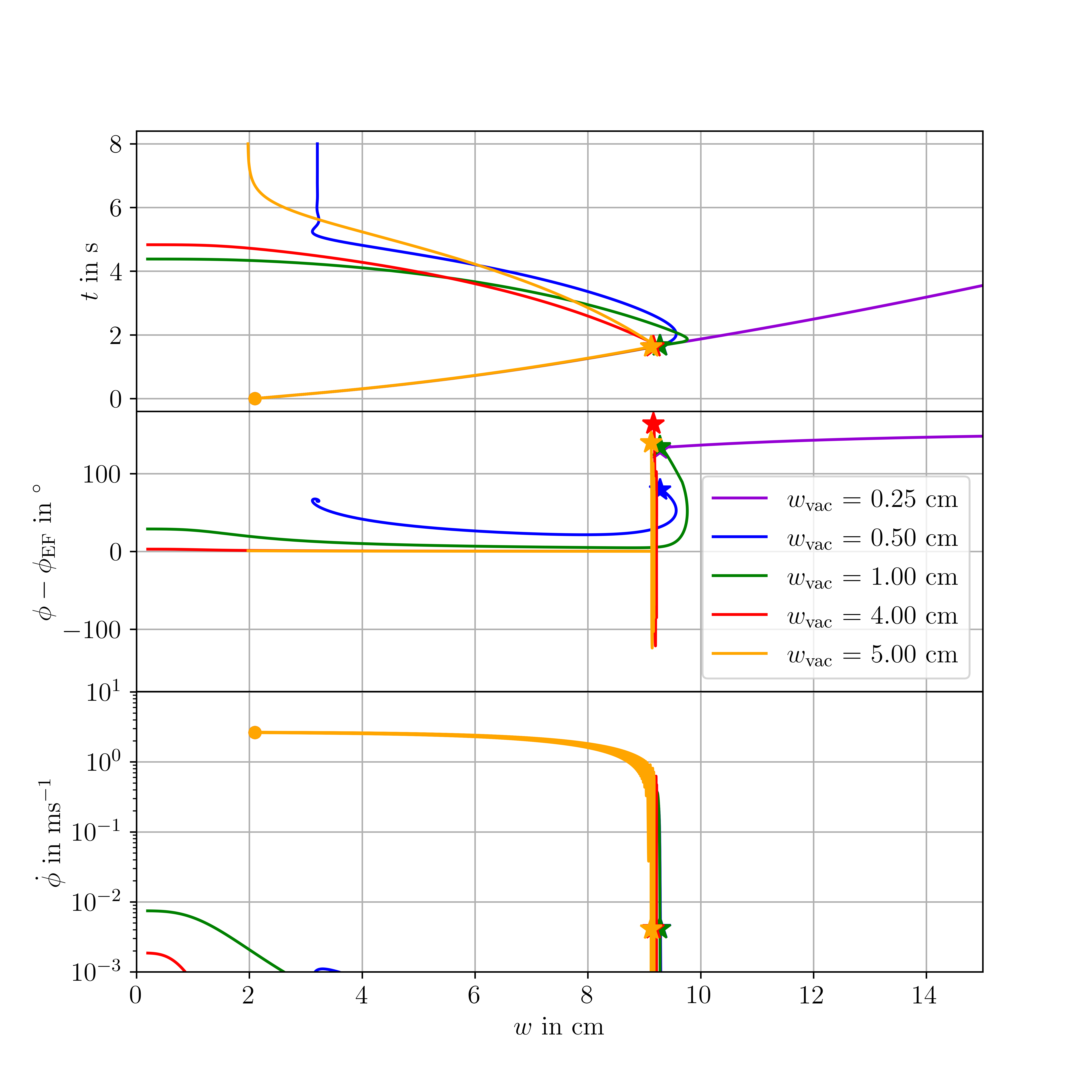}
    \caption{Island evolution for various vacuum island widths, $w_\mathrm{vac}$. Top: the island width evolution in time. (All of the curves start at $t=0$, and they initially overlap.)  Middle: The stars indicate the `locking' events, which is where the corresponding curves start.  Each of the curves show the value of the phase as a function of the island width $w$ following this event.  Bottom: the rate of change of the instantaneous rotation frequency as a function of the island width $w$. The large dot indicates the initial seeding event. The initial seeding width and rotation frequency is taken to be the same for each of the cases, and the curves initially overlap.}
    \label{fig:plot_phase_evol_w_vac}
\end{figure}

As $w_\mathrm{vac}^\mathrm{place}(w_\mathrm{marg}) < w_\mathrm{vac}^\mathrm{power}$, there exists a range of error fields for which locked island stabilisation can be employed to efficiently stabilise the $2/1$-NTM in ITER. These predictions are verified in Fig.~\ref{fig:plot_phase_evol_w_vac}, where locked island stabilisation with $5$~MW of power is simulated by evolving \eqref{eq:rutherford_ITER} and \eqref{eq:rotation_ITER} for different values of $w_\mathrm{vac}$. The broadening factor, predicted to be between $2.5$ and $3.5$ in ITER \cite{snicker_effect_2018}, was here assumed to be $3$.

As shown in the bottom panel of  Fig.~\ref{fig:plot_phase_evol_w_vac}, the island rotation in each of the cases is initially slowed by the interaction with the resistive wall.  As the islands slow, the error field exerts a sinusoidal force, slowing the rotation during part of the orbit and accelerating the rotation during part.  This leads to a small sinusoidal variation in the instantaneous rotation frequency that is superimposed on the average rotation frequency, with the island rotation in each period slowest when the island has a phase of 180 degrees relative to that of the error field.  We consider the island to `lock' when the instantaneous rotation frequency of the island falls below 0.4 Hz. Those locking events are indicated by stars in the middle panel of  Fig.~\ref{fig:plot_phase_evol_w_vac}, and the middle panel shows the subsequent evolution of the phase for each of the cases.  

For small $w_\mathrm{vac} = 0.25\;\mathrm{cm} < w_\mathrm{vac}^\mathrm{place}(w_\mathrm{lock})$, the error field is too small to lock the island at the desired phase, such that the locked island cannot be stabilised by the applied EC power and it grows instead. For $w_\mathrm{vac}^\mathrm{place}(w_\mathrm{lock}) < w_\mathrm{vac} = 0.50\;\mathrm{cm} < w_\mathrm{vac}^\mathrm{place}(w_\mathrm{marg})$, the error field is sufficiently large to initially lock the island at a small $\phi-\phi_\mathrm{EF}$ where it can be stabilised, but the island moves to larger $\phi-\phi_\mathrm{EF}$ when it reaches small island widths, preventing full stabilisation of the island. In the range $w_\mathrm{vac}^\mathrm{place}(w_\mathrm{marg}) < w_\mathrm{vac} = 1\;\mathrm{cm},\; 4\;\mathrm{cm} < w_\mathrm{vac}^\mathrm{power}$, the island is locked at small $\phi-\phi_\mathrm{EF}$ at all times, enabling full stabilisation. Finally, for $ w_\mathrm{vac} = 5\;\mathrm{cm} > w_\mathrm{vac}^\mathrm{power}$, the island can only be partially stabilised as the large error field increases the power required to stabilise the NTM \eqref{eq:criterion_P_RF_LM_large_wdep}.

The need to avoid born-locked-modes places a further restriction on the tolerable error field. In ITER, this leads to the 3-mode requirement $b_\mathrm{3-mode} = \sqrt{b_{r2,1}^2 + 0.8 b_{r3,1}^2 + 0.2 b_{r1,1}^2} \leq 5\cdot10^{-5}$ \cite{hender_chapter_2007} where $b_{rm,n}$ are radial components of the magnetic field with poloidal (toroidal) mode number $m$ ($n$), normalised by the toroidal field $B_t$. These can be related to the vacuum island width by assuming the current at $r>r_s$ is negligible, such that we can approximate the fall-off of the flux perturbation $\psi_{m,n}$ as $\psi_{m,n}(r_s) \approx \psi_{m,n}(a) (r_s/a)^m$. Further approximating the poloidal field as $B_p \approx B_t q r_s / R_0$ and using $\psi_{2,1}(a) = b_{r2,1} B_t a/m$ \cite{yu_locking_2008}, the vacuum island width is then given by \cite{fitzpatrick_interaction_1993}
\begin{equation}
    w_\mathrm{vac} = 4 r_s \sqrt{\frac{\psi_{2,1}(r_s)}{L_q B_p}} = 4 r_s \sqrt{ \frac{a R_0}{L_q r_s} \left( \frac{r_s}{a} \right)^2 b_\mathrm{r2,1} } \approx \sqrt{\frac{b_{r2,1}}{5\cdot 10^{-5}}} \cdot 10\;\mathrm{cm}.
\end{equation}
The 3-mode criterion is thus similar in scale to the criterion \eqref{eq:criterion_w_vac_stab}. Which of the two criteria imposes a more stringent condition on $b_\mathrm{r2,1}$ will depend on the partition of the 3-mode field into its components. In any case, the range of vacuum island widths ($\approx 1$ to $4$ cm) for which locked island stabilisation can be performed efficiently does not conflict with the $3$-mode criterion.

Here, the radial field at the edge and resonant surface were related assuming the current to be negligible for $r_s > a$. Even though the $2/1$-NTM is close to the plasma edge, as $r_s \approx 0.8 a$, a more detailed treatment ought to consider the shielding by the large bootstrap current in the H-mode pedestal. This should relax the criterion on the error field size, i.e. a smaller vacuum island width would result from the same error field at the edge.

\section{Discussion}
\label{sec:discussion}

In this work, we investigated the possibility of a locked island stabilisation strategy for NTMs in tokamaks. Such a strategy would be particularly relevant for large tokamaks like ITER, where islands lock at a small size and large broadening of the stabilising EC wave is expected, making island stabilisation during the rotating phase challenging. In these devices, allowing the NTMs to lock could be benign in terms of the impact on plasma confinement and disruptivity, as the ratio of the width at locking to the plasma minor radius is small, although it may be necessary to suppress the island on the momentum time scale in order to avoid a transition to L-mode. After locking, these islands can be stabilised rapidly and efficiently, reducing both the peak and average power requirements for NTM stabilisation compared to the rotating island stabilisation schemes.

Furthermore, implementing a locked island stabilisation strategy in ITER could help reduce the disruptivity to achieve the allowed $1\%$ of disruptions per discharge \cite{strait_progress_2019}. With the currently envisioned rotating island stabilisation strategy, it may be difficult to reliably avoid mode locking, especially in the case of large seeding events, e.g. due to sawteeth causing helical core deformations \cite{igochine_conversion_2014}, predicted to be significant for ITER \cite{wingen_use_2018}. A locked island stabilisation strategy should thus be planned, at least as a back-up strategy for those cases where rotating island stabilisation fails. Even if the locked mode grows to a large size, it might still be possible to stabilise it and avoid a disruption by leveraging the current condensation effect \cite{reiman_suppression_2018}.

We showed that a range of error field values exists for which locked island stabilisation is practical in ITER. The error field in ITER will be controlled using external coils \cite{hender_chapter_2007, amoskov_optimization_2015}, such that a locked mode stabilisation strategy might merely require an adjustment of the phase of the residual error field, to lock the island in front of the EC wave launchers. In particular, the error field magnitude need not be increased above the 3-mode criterion already being applied, as was shown in section~\ref{sec:ITER_error_field_requirements}.

The present work could be extended upon in multiple ways. A rotating island could be partially stabilised with a small amount of power such that it locks at a smaller width, further reducing the confinement degradation and allowing for faster and more efficient stabilisation of the smaller locked island. More detailed theoretical modeling of locked island stabilisation could be performed using MHD codes, as previously considered in one study \cite{yu_locking_2008}. Experimentally, the impact of small locked neoclassical tearing modes on confinement and loss of H-mode should be investigated, as well as their position control and stabilisation using EC waves, extending the pioneering work on the DIII-D tokamak \cite{volpe_advanced_2009, volpe_avoiding_2015}. 

\section{Conclusions}
\label{conclusions}

It has been commonly thought that
NTMs in ITER must be stabilised during the rotating
phase and locked modes (LMs) must be avoided at all
cost. However, islands will lock at smaller widths as tokamaks grow larger and as the externally imposed torque
becomes smaller relative to the plasma inertia. Moreover, new information has emerged in recent years with
regard to broadening of EC deposition profiles \cite{decker_effect_2012,brookman_experimental_2017, chellai_millimeter-wave_2018, chellai_millimeter-wave_2019,brookman_resolving_2021}
predicted to be large in ITER \cite{snicker_effect_2018} and even more so in the
proposed DEMO reactor \cite{snicker_transport_2021}, with regard to the effect of
the test blanket modules in ITER on the island width at
locking \cite{la_haye_effect_2017}, with regard to the short term robustness of
the H mode against locking \cite{nelson_experimental_2020,volpe_avoiding_2015,volpe_private_2021}, and with regard to
the benign effects of small islands in RMP ELM suppression experiments \cite{evans_suppression_2004}. The calculations here, taking the
new information into account, now challenge the standard paradigm for NTM stabilisation, even for ITER,
suggesting that locked mode stabilisation will be more advantageous, if not critical. With previous modeling and experimental work having focused almost entirely on RM
stabilisation, this paradigm shift should motivate modeling and experimental studies focused on the LM stabilisation scenario for ITER. It will be critical to the success
of ITER to robustly stabilise NTMs while minimising the
impact on the fusion gain, Q, and while freeing sufficient
EC power for ITER's other needs.

The authors thank R.J. La Haye and F.A.G. Volpe for helpful discussions. Simulation results presented in this study were obtained on the PPPL research cluster. This work was supported by U.S. DOE DE-AC02-09CH11466 and DE-SC0016072.\\

\bibliographystyle{unsrt}
\bibliography{references}

\end{document}